\documentclass[12pt]{article}
\newcommand{\bsize}{\small}
\usepackage{times}
\usepackage{w-thm}
\usepackage{natbib}
\usepackage{authblk}

\newcommand{\bx}{{\bf x}}

\newcommand{\bmu}{\mbox{\boldmath $\mu$}}

\newcommand{\bxi}{\mbox{\boldmath $\xi$}}

\newcommand{\bgamma}{\mbox{\boldmath $\Gamma$}}

\newcommand{\btheta}{\mbox{\boldmath $\theta$}}

\newcommand{\blambda}{\mbox{\boldmath $\Lambda$}}

\theoremstyle{plain}

\theoremstyle{definition}

\usepackage[]{graphicx}
\chardef\bslash=`\\ 

\usepackage{amsmath,amssymb,amsthm}
\usepackage{float}
\usepackage{booktabs}
\usepackage{longtable}
\usepackage{array}
\usepackage{multirow}
\usepackage{wrapfig}
\usepackage{float}
\usepackage{colortbl}
\usepackage{pdflscape}
\usepackage{tabu}
\usepackage{threeparttable}
\usepackage{threeparttablex}
\usepackage[normalem]{ulem}
\usepackage{makecell}
\usepackage{xcolor}
\usepackage{type1cm,verbatim,hyperref}
\usepackage{graphicx}
\usepackage{latexsym}
\usepackage{lscape}
\usepackage{subcaption}
\usepackage{todonotes}
\usepackage{setspace,color,bbm,amsfonts}

\def\blambda{\mbox{\boldmath $\lambda$}}

\def\btheta{\mbox{\boldmath $\theta$}}

\def\bbeta{\mbox{\boldmath $\beta$}}

\def\bxi{\mbox{\boldmath $\xi$}}

\def\bgamma{\mbox{\boldmath $\gamma$}}
\def\bmu{\mbox{\boldmath $\mu$}}

\def\bp{{\bf p}}

\def\bx{{\bf x}}

\def\bX{{\bf X}}

\def\bSig\mathbf{\Sigma}

\begin{document}
\def\spacingset#1{\renewcommand{\baselinestretch}%
{#1}\small\normalsize} \spacingset{1}


\title{\bf A likelihood approach to proper analysis of secondary outcomes in matched case-control studies}
\author[1]{Shanshan Liu}
\author[1]{Guoqing Diao\thanks{
Contact gdiao@email.gwu.edu
}}
\affil[1]{Department of Biostatistics and Bioinformatics, The George Washington University}
\date{}
\maketitle

\begin{abstract}
Matched case-control studies are commonly employed in epidemiological research for their convenience and efficiency. Analysis of secondary outcomes can yield valuable insights into biological pathways and help identify genetic variants of importance. Naive analysis using standard statistical methods, such as least-squares regression for quantitative traits, can be misleading because they fail to account for unequal sampling induced by the case-control design and matching. In this paper, we propose novel statistical methods that appropriately reflect the study design and sampling scheme in the analysis of secondary outcome data. The new methods provide consistent estimation and accurate coverage probabilities for the confidence interval estimators. We demonstrate the advantages of the new methods through simulation studies and a real application with diabetes patients. R code implementing the proposed methods is publicly available.
\end{abstract}

\noindent%
\vfill

\spacingset{1.45} 

\newpage






\section{Introduction}
Case-control designs are commonly employed in biomedical studies,
particularly in studying risk factors for rare diseases or conditions \citep{schlesselman1982case,lash_modern_2021}.
Under this design, a sample of cases and a sample of controls are
sampled from the underlying population and potential risk factors are
then collected retrospectively. Compared to cohort studies, case-control
studies are cost-effective and a prospective odds ratio can be recovered \citep{breslow1980statistical,prentice1979logistic}.
Conventionally, the main interest of case-control studies focuses on
assessing the association between the primary outcome, case-control
status, and the covariates of interest. Recent years have witnessed an
increasing number of case-control studies in the genetic analysis of
complex human diseases and related traits \citep{thomas2010methods,balding2006tutorial}. Many case-control genetic
association studies collect extensive information on related
quantitative or qualitative traits in addition to the disease trait, which are
called secondary outcomes or secondary traits. Exploring the
association between genetic factors and these secondary traits has
received increasing attention in genetic epidemiological studies. Such
analyses of secondary trait data can provide valuable insight into
biological pathways, and the association analysis of the primary disease
outcome and the secondary traits can be of interest themselves \citep{lin_proper_2009,ghosh2013unified}. For example, the Diabetes Genetics Initiative (DGI) study recruited patients with type 2 diabetes and controls, while collecting data on
numerous secondary traits, including anthropometric measures, glucose
tolerance and insulin secretion, lipids, and blood
pressure \citep{diao_analysis_2018}. It is convenient and cost-effective to
study the association between secondary traits using data from existing
case-control studies.

As pointed out by \cite{lin_proper_2009}, naive
regression methods are biased when used for analyzing secondary traits due to the biased sampling inherent in the case-control design. To overcome the limitations of the aforementioned
naive methods, the authors developed efficient likelihood-based
procedures that properly reflect the case-control sampling for the
association analysis of secondary quantitative and qualitative traits in
case-control studies without distribution assumptions on other
covariates of interest. Other methods have also been proposed, including weighted estimating equations \citep{richardson2007analyses,monsees2009genome,song2016general,pan2018secondary,zhou2020analysis}, likelihood-based approaches \citep{ghosh2013unified,tchetgen_tchetgen_general_2014}, copula modeling \citep{he2012gaussian}, and semiparametric models \citep{wei2013robust,ma2016semiparametric}.

It is common for case-control studies to adopt a matched design, either
by individual matching or frequency matching \citep{breslow1980statistical,schlesselman1982case,lash_modern_2021}. An
example of the latter is selecting cases and controls in males and
females separately. The purpose of matching in case-control designs is to
increase statistical and cost efficiency \citep{rothman2008modern,pearce2016analysis}. When analyzing matched
case-control studies, the Mantel-Haenszel odds ratio estimator \citep{mantel1959statistical} or the
conditional logistic regression model \citep{cox1972regression} is needed to account for the
selection bias introduced during sampling. Since matching further adds to the
complexity of the analysis of secondary traits, the aforementioned methods cannot be applied directly in this setting. To our knowledge, there has been limited published methodological research specifically addressing this problem and no publicly available, user-friendly software currently exists to implement such approaches. Therefore, we propose a new method for the proper analysis of the association between secondary traits and risk factors in matched case-control studies by modifying the methods of \cite{lin_proper_2009} to account for the matching. The proposed method is implemented in a publicly available R program.  

The paper is organized as follows: in Section 2, we formulate the problem and describe the proposed estimator. In Section 3, extensive simulations are conducted in different settings to evaluate the performance of the proposed method relative to the naive methods. In Section 4, we illustrate applications of our method using a large dataset of  patients with diabetes.

\section{Methods}\label{methods}

We first introduce some notation. Let $D$ denote the disease status, which
takes value 1 for cases and 0 for controls. Let $Z$ denote the matching
strata taking values \(1,\dots,K\), which is a categorical variable or
derived from a combination of multiple categorical variables. Let $Y$
denote the secondary outcome and \(\mathbf{X}\) a vector of
covariates. The joint
distribution of a randomly sampled observation \((D, Z, Y, \bX)\) is
\[P(D, Z, Y, \bX) = P(D|Z,Y,\bX)P(Y|Z,\bX)P(\bX|Z)P(Z).\] The retrospective
likelihood for an observation from a matched case-control sample is
\[P(Y,\bX|D,Z) = \frac{P(D, Z, Y, \bX)}{P(D, Z)} = \frac{P(D|Z,Y,\bX)P(Y|Z,\bX)P(\bX|Z)}{\int_{\mathcal{X},\mathcal{Y}} P(D|Z,y,\bx)P(y|Z,\bx)P(\bx|Z) \, d \bx dy},\]
where $\mathcal{X}$, and $\mathcal{Y}$ are the supports of $\bX$ and $Y$, respectively. The likelihood reflects the fact that the sampling is conditional on both
the case-control status $D$ and the matching variable $Z$. We consider the
following logistic model for the conditional distribution of $D$ given $\bX$,
$Y$ and $Z$,
\[P(D=1|\bX,Y,Z=k) = \frac{\exp(\gamma_{0k}+\bgamma_1^T \bX+\gamma_2Y)}{1+\exp(\gamma_{0k}+\bgamma_1^T \bX+\gamma_2Y)}.\]
In the secondary outcome analysis, our main interest is the effect of
$\bX$ on $Y$, assuming the effect is homogeneous across strata of $Z$. For a binary $Y$,
we consider the following logistic model for the conditional
distribution of $Y$ given $\bX$ and $Z$,
\[P(Y=1|\bX,Z=k) = \frac{\exp(\beta_{0k}+\bbeta_1^T \bX)}{1+\exp(\beta_{0k}+\bbeta_1^T \bX)}.\]
For a continuous $Y$, we use the linear regression model, specifying the
conditional mean of $Y$ as \(\beta_{0k}+\bbeta_1^T \bX\) and variance
\(\sigma^2\). We focus on binary $Y$ hereon.

For the conditional distribution of \(\mathbf{X}\) given $Z$, we take a
nonparametric approach. In particular, let \(P(\bX_{ki}|Z=k)=p_{ki}\),
where \(\bX_{ki}\) is the vector covariates for the \(i\)th subject in the
\(k\)th stratum. Denote the total number of observations in the \(k\)th
stratum by \(n_k\). We assume that observations are sampled independently within
each stratum and that the strata are independent. Define $\bgamma = (\gamma_{01},...,\gamma_{0K}, \bgamma_1, \gamma_2)$, $\bbeta=(\beta_{01},...,\beta_{0K}, \bbeta_1)$, and $\bp=(p_{11},...,p_{1n_1},...,p_{K1},...,p_{Kn_K})$. The probabilities $p_{ki} ~(k=1,...,K; i=1,\dots,n_k)$ satisfy the constraint  \(\sum_{i=1}^{n_k} p_{ki} = 1\). Then the likelihood
function for the unknown parameters $\btheta=(\bgamma, \bbeta, \bp)$ based on the observations in the entire sample is
\[
L(\btheta) = \prod_{k=1}^K \prod_{i=1}^{n_k} \frac{ 
\frac{\exp\{D_{ki} (\gamma_{0k}+\bgamma_1^T \bX_{ki}+\gamma_2 Y_{ki})\}}{1+\exp(\gamma_{0k}+\bgamma_1^T \bX_{ki}+\gamma_2 Y_{ki})}
P(Y_{ki}|\bX_{ki})
p_{ki}}{
\sum_{j=1}^{n_k}\int_{\mathcal{Y}}
\frac{\exp\{D_{ki} (\gamma_{0k}+\bgamma_1^T \bX_{kj} +\gamma_2 y)\}}{1+\exp(\gamma_{0k}+\bgamma_1^T \bX_{kj} +\gamma_2 y)}
P(y|\bX_{kj})
p_{kj}\, dy
}.\] 
Throughout, we focus on stratified case–control designs with non-negligible stratum sizes, so that within-stratum quantities are estimable. Individually matched designs fall outside this framework. We consider three scenarios: (1) rare disease for each category of $Z$; (2)
the disease rate in each category of $Z$ is known; and (3) the disease
rates in all categories of $Z$ are unknown. These different scenarios lead to either simplification of the likelihood function or additional constraints on the unknown parameters. 
The likelihood function above has a form similar to that often encountered in the empirical likelihood approach and can be maximized using the same techniques. Details on this connection are discussed in the Appendix. In short, we first derive a profile likelihood function of $(\bgamma,\bbeta)$ and then use an optimization algorithm to maximize the profile likelihood. We can show that the resulting maximum likeihood estimators are consistent, asymptotically normal, and the variance-covariance estimators attain the semiparametric efficiency bound. The theoretical and computational details are provided in the Appendix.

\section{Simulation Study}\label{simulation-study}

We conducted simulation studies to assess the performance of
naive conditional and unconditional regression models, and the new
methods in the analysis of secondary quantitative traits.

We consider a binary disease outcome $D$, a binary matching factor $Z$, a
binary secondary outcome $Y$, and a continuous (standard normal)
covariate $X$. We are interested in measuring the association between $X$
and $Y$. The disease model is a logistic regression model
\( \text{logit} P(D=1|X,Y,Z=k) = \gamma_{0k} + \gamma_1 X + \gamma_2 Y\).
We set \(\gamma_1 = \log(0.5), \gamma_2 = \log(0.1)\), and choose the
values of \(\gamma_{01}\) and \(\gamma_{02}\) to yield a disease rate of
1\%, 5\%, and 10\% in the entire population. The secondary outcome model is a logistic
regression model
\( \text{logit} P(Y=1|X,Z=k) = \beta_{0k} + \beta_1 X \).
We set \(\beta_{01} = -1, \beta_{02} = -0.2\), and \(\beta_1 = \log 2\). The choice of \(\beta_1 = \log 2\) represents a moderate association
between the covariate and the secondary outcome.

We generate a source population of 2 million, 1.2 million in stratum 1
($Z=1$), and 0.8 million in stratum 2 ($Z=2$). For each level of disease
rate, we generated 10000 data sets with $n=500$ cases and
controls per stratum with a total of 2000, or 1000 cases and
controls per stratum with a total of 4000. 
We compare the following naive methods:
(1) Conditional logistic regression of $Y$ on $X$, conditional on the
combination of $Z$ and $D$ as stratum (Conditional), (2) Unconditional,
unadjusted logistic regression of $Y$ on $X$ (``Unconditional, unadjusted"),
(3) Unconditional logistic regression of $Y$ on $X$, adjusted for $Z$
(``Unconditional, adjusted 1"), and (4) Unconditional logistic regression
of $Y$ on $X$, adjusted for $D$ and $Z$ (``Unconditional, adjusted 2"), to the
following proposed methods: (1) assuming rare disease (PM1), (2)
assuming disease rates are known and not rare (PM2), and (3)
assuming disease rates are unknown and not rare (PM3).

We summarize the results in Table 1 and Table 2 for $n=500$ and 1000, respectively, showing the biases of the effect estimates and the coverage probabilities of 95\% confidence
intervals for \(\beta_1\). None of the naive regression methods consistently performs well in all scenarios. The conditional regression model fails to converge when $n=1000$ under all scenarios, but is only slightly biased and produces MSE and coverage probabilities comparable to the proposed methods when $n=500$. The ``unconditional, unadjusted" and ``unconditional, adjusted 1" models both produce biased estimates and suffer from serious undercoverage issues in all scenarios. The ``unconditional, adjusted 2" model produces unbiased estimates and confidence intervals at the nominal level when disease is rare at 1\%, but the biases grow and coverage probabilities drop below the nominal level when disease rates are 5\%, and 10\%. In contrast, the new methods perform better than all 
naive regression methods with little bias and variance estimates correctly
reflect the true variations. When the true disease rate is
low (around 1\%), PM1 (assuming rare disease) has the lowest MSE
as the simplification approximates the true model very well. PM2
(known disease rates) has a slightly smaller bias as a benefit of
incorporating external information, but this advantage is counteracted by
the increased variance caused by additional model complexity. PM3
(estimating disease rates) has a similar performance to PM1 but is
unstable when the sample size is moderate ($n=500$) because with an extremely
low disease rate, the number of diseased observations is small, making
it challenging to estimate the disease rates. When the disease is not
rare (disease rate 5-10\%), PM2 performs the best in most settings,
as expected. The performance gain is due to the additional information,
i.e., disease rates, incorporated in the estimation process. PM3
also has good performance and is the best when disease rates are
unknown. PM1 has a larger bias than the other two new methods
because the rare disease approximation is no longer valid. As a result, the 95\% confidence
intervals all have proper coverage probabilities, whereas the estimates from the naive regressions are biased, resulting in undercoverage of the confidence intervals. The conditional logistic regression models
would fail to converge with large sample sizes, which is a known issue
with this model.

\begin{table}[!h]
\centering\centering
\begin{threeparttable}
\caption{\label{tab:unnamed-chunk-3}Summary statistics for conditional, unconditional, and new methods based on 10000 replicates for sample size $n=500$}
\centering
\fontsize{12}{14}\selectfont
\begin{tabular}[t]{lrrrrrr}
\toprule
 & $\text{Bias}$ & RB & Mean SE & EmpSD & MSE & CP\\
\midrule
\addlinespace[0.3em]
\multicolumn{7}{l}{{$p$=1\%}}\\
\hspace{1em}Conditional & -0.010 & -0.015 & 0.070 & 0.068 & 0.476 & 0.95\\
\hspace{1em}Unconditional,unadjusted & 0.234 & 0.338 & 0.064 & 0.062 & 5.842 & 0.03\\
\hspace{1em}Unconditional,adjusted1 & 0.260 & 0.376 & 0.065 & 0.063 & 7.173 & 0.01\\
\hspace{1em}Unconditional,adjusted2 & -0.009 & -0.012 & 0.070 & 0.068 & 0.475 & 0.95\\
\hspace{1em}PM1 & -0.008 & -0.012 & 0.069 & 0.068 & 0.464 & 0.95\\
\hspace{1em}PM2 & -0.006 & -0.008 & 0.069 & 0.069 & 0.480 & 0.95\\
\hspace{1em}PM3 & 0.034 & 0.049 & 0.069 & 0.123 & 1.627 & 0.94\\
\addlinespace[0.3em]
\multicolumn{7}{l}{{$p$=5\%}}\\
\hspace{1em}Conditional & -0.026 & -0.038 & 0.070 & 0.069 & 0.547 & 0.93\\
\hspace{1em}Unconditional,unadjusted & 0.230 & 0.333 & 0.064 & 0.064 & 5.712 & 0.04\\
\hspace{1em}Unconditional,adjusted1 & 0.256 & 0.371 & 0.065 & 0.065 & 6.994 & 0.01\\
\hspace{1em}Unconditional,adjusted2 & -0.025 & -0.036 & 0.070 & 0.069 & 0.543 & 0.93\\
\hspace{1em}PM1 & -0.024 & -0.035 & 0.069 & 0.069 & 0.530 & 0.93\\
\hspace{1em}PM2 & -0.008 & -0.012 & 0.069 & 0.069 & 0.479 & 0.95\\
\hspace{1em}PM3 & -0.003 & -0.005 & 0.069 & 0.070 & 0.495 & 0.95\\
\addlinespace[0.3em]
\multicolumn{7}{l}{{$p$=10\%}}\\
\hspace{1em}Conditional & -0.039 & -0.056 & 0.069 & 0.070 & 0.643 & 0.90\\
\hspace{1em}Unconditional,unadjusted & 0.219 & 0.317 & 0.063 & 0.064 & 5.212 & 0.06\\
\hspace{1em}Unconditional,adjusted1 & 0.246 & 0.355 & 0.065 & 0.065 & 6.456 & 0.03\\
\hspace{1em}Unconditional,adjusted2 & -0.037 & -0.054 & 0.070 & 0.070 & 0.636 & 0.90\\
\hspace{1em}PM1 & -0.036 & -0.053 & 0.069 & 0.070 & 0.621 & 0.91\\
\hspace{1em}PM2 & -0.005 & -0.007 & 0.068 & 0.069 & 0.476 & 0.94\\
\hspace{1em}PM3 & -0.015 & -0.022 & 0.069 & 0.073 & 0.553 & 0.93\\
\bottomrule
\end{tabular}
\begin{tablenotes}[flushleft]
\item \textit{Note:} Bias is the difference between the sample mean of the estimates and the true parameter value.
RB is relative bias. Mean SE is the sample mean of the standard error estimates. EmpSD is the sample standard deviation of the estimates.
MSE is the mean squared error multiplied by 100. CP is the empirical coverage probability of the 95\% confidence interval. 
\end{tablenotes}
\end{threeparttable}
\end{table}

\begin{table}[!h]
\centering\centering
\begin{threeparttable}
\caption{\label{tab:unnamed-chunk-3}Summary statistics for conditional, unconditional, and new methods based on 10000 replicates for sample size $n$=1000}
\centering
\fontsize{12}{14}\selectfont
\begin{tabular}[t]{lrrrrrr}
\toprule
 & $\text{Bias}$ & RB & Mean SE & EmpSD & MSE & CP\\
\midrule
\addlinespace[0.3em]
\multicolumn{7}{l}{{$p$=1\%}}\\
\hspace{1em}Conditional & NA & NA & 0.000 & NA & NA & \vphantom{2} NA\\
\hspace{1em}Unconditional,unadjusted & 0.233 & 0.336 & 0.045 & 0.042 & 5.596 & 0.00\\
\hspace{1em}Unconditional,adjusted1 & 0.259 & 0.375 & 0.046 & 0.043 & 6.896 & 0.00\\
\hspace{1em}Unconditional,adjusted2 & -0.010 & -0.014 & 0.049 & 0.047 & 0.232 & 0.95\\
\hspace{1em}PM1 & -0.009 & -0.014 & 0.049 & 0.046 & 0.225 & 0.95\\
\hspace{1em}PM2 & -0.007 & -0.010 & 0.049 & 0.052 & 0.277 & 0.94\\
\hspace{1em}PM3 & 0.013 & 0.018 & 0.049 & 0.047 & 0.238 & 0.95\\
\addlinespace[0.3em]
\multicolumn{7}{l}{{$p$=5\%}}\\
\hspace{1em}Conditional & NA & NA & 0.000 & NA & NA & \vphantom{1} NA\\
\hspace{1em}Unconditional,unadjusted & 0.228 & 0.330 & 0.045 & 0.044 & 5.404 & 0.00\\
\hspace{1em}Unconditional,adjusted1 & 0.254 & 0.367 & 0.046 & 0.045 & 6.648 & 0.00\\
\hspace{1em}Unconditional,adjusted2 & -0.027 & -0.039 & 0.049 & 0.048 & 0.306 & 0.92\\
\hspace{1em}PM1 & -0.027 & -0.038 & 0.049 & 0.048 & 0.298 & 0.92\\
\hspace{1em}PM2 & -0.011 & -0.016 & 0.048 & 0.048 & 0.247 & 0.94\\
\hspace{1em}PM3 & -0.004 & -0.005 & 0.049 & 0.049 & 0.240 & 0.95\\
\addlinespace[0.3em]
\multicolumn{7}{l}{{$p$=10\%}}\\
\hspace{1em}Conditional & NA & NA & 0.000 & NA & NA & NA\\
\hspace{1em}Unconditional,unadjusted & 0.218 & 0.315 & 0.045 & 0.045 & 4.942 & 0.00\\
\hspace{1em}Unconditional,adjusted1 & 0.244 & 0.352 & 0.046 & 0.046 & 6.146 & 0.00\\
\hspace{1em}Unconditional,adjusted2 & -0.040 & -0.057 & 0.049 & 0.049 & 0.399 & 0.86\\
\hspace{1em}PM1 & -0.039 & -0.056 & 0.049 & 0.049 & 0.386 & 0.87\\
\hspace{1em}PM2 & -0.007 & -0.010 & 0.048 & 0.048 & 0.236 & 0.94\\
\hspace{1em}PM3 & -0.016 & -0.023 & 0.049 & 0.050 & 0.272 & 0.93\\
\bottomrule
\end{tabular}
\begin{tablenotes}[flushleft]
\item \textit{Note:} Bias is the difference between the sample mean of the estimates and the true parameter value.
RB is relative bias. Mean SE is the sample mean of the standard error estimates. EmpSD is the sample standard deviation of the estimates.
MSE is the mean squared error multiplied by 100. CP is the empirical coverage probability of the 95\% confidence interval. 
\end{tablenotes}
\end{threeparttable}
\end{table}

\section{Real Data Analysis}\label{real-data-analysis}

We apply the proposed methods to the Diabetes 130-US Hospitals for Years 1999-2008 dataset from
the UCI Machine Learning Repository \citep{diabetes_130}. It contains 101766 admissions of
diabetes patients. We consider a hypothetical matched case-control study
with 30-day readmission as the disease outcome, and age (below/above 50 years)
and gender as matching factors ($Z_1$ and $Z_2$). We are interested in the association
between $X$, the number of diagnoses at admission (greater than 3), and $Y$,
the length of stay (longer than 3 days).

We draw random samples of 50, 200, and 750 cases and controls in each
stratum 1000 times. We compare the following naive models with the
new methods. The difference between the last two models is in whether
the interactions between matching factors are included in the model.\\
(i) Conditional: conditional on age-gender-disease stratum\\
(ii) Unconditional, unadjusted: standard logistic regression of $Y \sim X$\\
(iii) Unconditional, adjusted1: standard logistic regression of $Y \sim X+Z_1+Z_2$\\
(iv) Unconditional, adjusted2: standard logistic regression of
$Y \sim X+Z_1+Z_2+D$\\
(v) Unconditional, adjusted3: standard logistic regression of
$Y \sim X+Z_1+Z_2+Z_1\times Z_2+D$

Table 3 summarizes the results of the five naive methods described above and the proposed methods. The biases are calculated relative to the beta estimate obtained using all 101766 diabetes patients in the dataset. 
The conditional regression model fails to converge when stratum sizes are large, and produces large variance when sample size is small. The four unconditional regression models produce moderately biased estimates. Although the biases decrease as more covariates are included, they are not completely removed. PM1 does not perform well as expected because the disease
outcome is not rare (\textgreater10\%). But PM2 and PM3 always have
smaller biases and standard errors than the naive methods. It should be noted that when the sample size is small, the standard logistic
regressions may have large variances. This is a common issue when the
number of outcome events is small and unbalanced among covariate
levels. The new methods, however, seem to provide stable estimates under
these circumstances. 

\begin{table}[!h]
\centering\centering
\begin{threeparttable}
\caption{\label{tab:unnamed-chunk-4}Summary statistics for naive and new methods in analysis of diabetes data based on 1000 replicates}
\centering
\fontsize{12}{14}\selectfont
\begin{tabular}[t]{lrrr}
\toprule
 & Bias & RB & Mean SE\\
\midrule
\addlinespace[0.3em]
\multicolumn{4}{l}{\textbf{$n=50$}}\\
\hspace{1em}Conditional & 0.245 & 0.217 & 10.870\\
\hspace{1em}Unconditional,unadjusted & 0.381 & 0.337 & 2.424\\
\hspace{1em}Unconditional,adjusted1 & 0.284 & 0.251 & 2.422\\
\hspace{1em}Unconditional,adjusted2 & 0.252 & 0.223 & 2.410\\
\hspace{1em}Unconditional,adjusted3 & 0.262 & 0.231 & 2.408\\
\hspace{1em}PM1 & 0.213 & 0.188 & 0.581\\
\hspace{1em}PM2 & 0.208 & 0.184 & 0.564\\
\hspace{1em}PM3 & 0.231 & 0.204 & 0.644\\
\addlinespace[0.3em]
\multicolumn{4}{l}{\textbf{$n=200$}}\\
\hspace{1em}Conditional & 0.097 & 0.086 & 0.260\\
\hspace{1em}Unconditional,unadjusted & 0.240 & 0.212 & 0.256\\
\hspace{1em}Unconditional,adjusted1 & 0.141 & 0.125 & 0.258\\
\hspace{1em}Unconditional,adjusted2 & 0.100 & 0.088 & 0.260\\
\hspace{1em}Unconditional,adjusted3 & 0.102 & 0.090 & 0.260\\
\hspace{1em}PM1 & 0.102 & 0.090 & 0.270\\
\hspace{1em}PM2 & 0.096 & 0.085 & 0.260\\
\hspace{1em}PM3 & 0.100 & 0.089 & 0.260\\
\addlinespace[0.3em]
\multicolumn{4}{l}{\textbf{$n=750$}}\\
\hspace{1em}Conditional & NA & NA & NA\\
\hspace{1em}Unconditional,unadjusted & 0.233 & 0.206 & 0.131\\
\hspace{1em}Unconditional,adjusted1 & 0.136 & 0.120 & 0.132\\
\hspace{1em}Unconditional,adjusted2 & 0.093 & 0.083 & 0.132\\
\hspace{1em}Unconditional,adjusted3 & 0.095 & 0.084 & 0.132\\
\hspace{1em}PM1 & 0.094 & 0.083 & 0.137\\
\hspace{1em}PM2 & 0.093 & 0.082 & 0.132\\
\hspace{1em}PM3 & 0.084 & 0.074 & 0.133\\
\bottomrule
\end{tabular}
\begin{tablenotes}[flushleft]
\footnotesize
\item \textit{Note:} RB is relative bias. 
\end{tablenotes}
\end{threeparttable}
\end{table}

\section{Discussion}\label{discussion}
{\setlength{\parskip}{0pt}%
We develop semiparametric methods for the appropriate analysis of secondary outcomes in matched case-control studies. The three new methods outperform the naive methods in all
the settings in the simulation studies corresponding to the three scenarios for the disease rate.

In general, the naive regression methods perform worse than
the new methods in all simulation settings under examination.
Conditional regression models fail to converge when the sample size is
large. This is a known issue with computation involving conditional
likelihood. A recursion method has been proposed \citep{gail_likelihood_1981},
but the calculation is still prohibitive with very large sample sizes \citep{therneau_package_2023}. Various approximation methods are available but they
are not implemented here as it is out of scope. The unadjusted and
inadequately adjusted regression models produce seriously biased
parameter estimates. The conditional regression models and fully
adjusted regression models occasionally produce results comparable to
the new methods, but these are highly dependent on the strength and
specific form of the association between the secondary outcome and the
disease status. Admittedly, the proposed methods are more computationally intensive than the naive methods; however, the runtime remains tractable in practice. In real data analysis, it took 38 minutes to run 1000 simulations with a sample size of 3000 (750 in each stratum) on a MacBook Pro with an Apple M4 processor, i.e., about 2 seconds to analyze one data set.
We provide a few recommendations based on the numerical studies for the analysis of secondary outcomes in matched case-control studies:\par
(i) When the disease is rare, PM1 (assuming rare disease) is the
preferred approach. In this setting, PM3 (estimating disease rates)
can be unstable with small to moderate sample sizes, but can be
considered when the sample size is large. PM2 (known disease rates)
does not show significant advantages to justify the computational
complexity.\par
(ii) When the disease is not rare and the disease rate in each matching stratum
is known, PM2 (known disease rates) is the recommended approach.
Incorporating external information reduces bias and improves
efficiency.\par
(iii) When the disease is not rare and the disease rate in each matching stratum
is unknown, PM3 (estimating disease rates) is the recommended
approach. This method works well when the sample size is at least moderately large, 
but it can become unstable when the sample size is too small, in which case
PM1 may be considered. Above all, it is always better to obtain disease rates and use PM2.\par
(iv) Among naive regression methods, conditional regression
(conditioning on both the matching factors and the disease status) and
standard regression adjusting for both the matching factors and the disease
status are less biased and have lower MSEs than the other methods. But
both have limitations: conditional regression models may not converge
with large sample sizes; and their performance depends on the strength and
specific form of the association between the secondary outcome and the
disease. They are not recommended for principled analysis, but can be
used to generate sensible initial values for the new methods.

The proposed methods are developed for stratified case–control designs in which the matching variables define a finite number of discrete strata whose sizes increase with the total sample size. Accordingly, the framework does not apply directly when matching is performed on a continuous variable or under extremely fine matching that yields many small strata. In practice, however, it is common to coarsen continuous matching variables into clinically or substantively meaningful categories prior to analysis, as is routinely done in stratified and conditional modeling approaches (e.g., Mantel–Haenszel estimation). The proposed methods may then be applied to the resulting discrete strata.

An alternative approach is to estimate stratum-specific parameters, and then combine them for common parameters across strata. Examples include the split-and-conquer methods \citep{chen2014split} and the distributed estimation methods \citep{battey2018distributed,jordan2019communication,slud2018combining}. These estimators have a similar form to that of the fixed-effects meta-analyses \citep{lin2010relative,zeng2015random}. These estimators have been shown to be asymptotically equivalent to the joint estimator, but may lose efficiency in finite samples \citep{chen2020relative}. After combining the common parameter estimates, the stratum-specific estimates can be updated once more and used for prediction. Similar strategies have been adopted in meta-learning and transfer learning \citep{finn2017model,zhuang2020comprehensive}. Future work could explore finite-sample performance of these methods in comparison to the joint estimators.

\vspace{5mm}
\noindent {\bf{Conflict of Interest}}

\noindent {\it{The authors have declared no conflict of interest. }}

\vspace{5mm}
\noindent {\bf Supplementary Materials}

\noindent {\it R code for the real data analysis is available.}

\vspace{5mm}
\noindent {\bf {Data Availability}}

\noindent {\it The data used for the real application are publicly available at \url{https://archive.ics.uci.edu/dataset/296/diabetes+130-us+hospitals+for+years+1999-2008}.}

\section*{Appendix}

\subsection*{A.1.\enspace Maximum Likelihood Estimation}\label{i.-likelihood-equations}
\vspace*{12pt}

Given the observed data, the likelihood function of the unknown parameters $\btheta$ is given by
\[
\begin{split}
 L(\btheta) &= \prod_{k=1}^{K} \prod_{i=1}^{n_k} \left\{ \frac{p_{ki} P(Y_{ki}|\bX_{ki}) e^{\gamma_{0k} + \bgamma_1^T \bX_{ki}+ \gamma_2 Y_{ki}}/(1+ e^{\gamma_{0k} + \bgamma_1^T \bX_{ki}+ \gamma_2 Y_{ki}})}{P(D_{ki}=1)} \right\}^{D_{ki}} \\
 & \times \prod_{k=1}^{K} \prod_{i=1}^{n_k} \left\{ \frac{p_{ki} P(Y_{ki}|\bX_{ki}) /(1+ e^{\gamma_{0k} + \bgamma_1^T \bX_{ki}+ \gamma_2 Y_{ki}})}{P(D_{ki}=0)} \right\}^{1-D_{ki}},
 \end{split}
\]
where
\[
P(D_{ki}=1) = \sum_{j=1}^{n_k} \int_{\mathcal{Y}}  p_{kj} P(y|\bX_{kj}) e^{\gamma_{0k} + \bgamma_1^T \bX_{ki}+ \gamma_2 y}/(1+ e^{\gamma_{0k} + \bgamma_1^T \bX_{ki}+ \gamma_2 y})\, dy, 
\]
\[
P(D_{ki}=0) = \sum_{j=1}^{n_k} \int_{\mathcal{Y}}  p_{kj} P(y|\bX_{kj})/(1+ e^{\gamma_{0k} + \bgamma_1^T \bX_{ki}+ \gamma_2 y} )\, dy,
\]
and \(p_{ki} = P(\bX_{ki}|Z=k)\) was previously defined in Section 2. The model for the secondary outcome $Y$ given covariates $\bX$ can be a specified parametric
model, e.g., logistic regression for binary $Y$ and linear regression for
continuous $Y$. In the case of binary $Y$, the integration in the following
derivation becomes a summation over 0 and 1. We next show how to derive the profile likelihood function of $(\bbeta, \bgamma)$, i.e., $l_p(\bbeta, \bgamma) = \max_{\bp \in \mathcal{P}} \log L(\btheta)$, where $\mathcal{P}$ is the parameter space for $\bp$ that satisfies the aforementioned constraints. 

We first consider the scenario with a rare disease. Under the rare disease assumption, it can be shown that 
\[
P(D=1|\bX, Z=k, Y) \approx e^{\gamma_{0k} + \bgamma_1^T \bX+ \gamma_2 Y} ~~
\text{and} ~~ P(D=0|\bX, Z=k, Y) \approx 1.
\]
The likelihood function can then be simplified as 
\[
\begin{split}
L(\btheta) & \approx 
\prod_{k=1}^{K} \prod_{i=1}^{n_k} \left\{ \frac{p_{ki} P(Y_{ki}|\bX_{ki})e^{\bgamma_1^T \bX_{ki}+\gamma_2 Y_{ki}}}{\sum_{j=1}^{n_k} \int_{\mathcal{Y}} p_{kj} P(y|\bX_{kj})  e^{\bgamma_1^T \bX_{ki}+\gamma_2 y} \, dy} \right\}^{D_{ki}} \times \prod_{k=1}^{K} \prod_{i=1}^{n_k} \{p_{ki} P(Y_{ki}|\bX_{ki})\}^{1-D_{ki}}\\
& = \prod_{k=1}^{K} \prod_{i=1}^{n_k} p_{ki} P(Y_{ki}|\bX_{ki}) \left\{ \frac{e^{\bgamma_1^T \bX_{ki}+\gamma_2 Y_{ki}}}{\sum_{j=1}^{n_k} \int_{\mathcal{Y}} p_{kj} P(y|\bX_{kj})  e^{\bgamma_1^T \bX_{ki}+\gamma_2 y} \, dy} \right\}^{D_{ki}}.\\
\end{split}
\]

Recall that $p_{ki}$s satisfy the constraints 
\(\sum_{i=1}^{n_k} p_{ki} =1\) for \(k=1, ..., K.\) 
Using the Lagrange multiplier, we differentiate the log likelihood $l(\btheta) = \log L(\btheta)$ with
respect to \(p_{ki}\) subject to the constraints and set the derivative to be 0,
\[
\frac{\partial \{l(\btheta) -  \sum_{k=1}^K\lambda_k (\sum_{i=1}^{n_k} p_{ki} -1) \}}{\partial{p_{ki}}}  = \frac{1}{p_{ki}} - n_{1k} \frac{ \int_{\mathcal{Y}} P(y|\bX_{ki}) e^{\bgamma_1^T \bX_{ki} + \gamma_2 y} \, dy}{\sum_{j=1}^{n_k} \int_{\mathcal{Y}} P(y|\bX_{kj}) e^{\bgamma_1^T \bX_{kj}+\gamma_2 y} dy } - \lambda_k=0,
\]
where $n_{1k} = \sum_{i=1}^{n_k} D_{ki}$ is the number of cases in the $k$th stratum.
Multiplying the above equation by \(p_{ki}\) and summing over $i=1,...,n_k$, we have
\(\lambda_k = n_k - n_{1k}\) for \(k=1, ..., K\).
Plugging \(\lambda_k = n_k - n_{1k}\) into the above equation, we obtain 
\[
p_{ki} = \left\{ n_k - n_{1k} + n_{1k} \xi_k \int_{\mathcal{Y}} P(y|\bX_{ki})e^{\bgamma_1^T \bX_{ki}+ \gamma_2 y} \,dy \right\}^{-1},
\]
where
\(\xi_k = \{ \sum_{j=1}^{n_k} \int_{\mathcal{Y}} p_{kj} P(y|\bX_{kj}) e^{\bgamma_1^T \bX_{kj} + \gamma_2 y} dy\}^{-1}\).

The profile log-likelihood function for $(\bbeta, \bgamma)$ then takes the form
\[
\begin{split}
l_{p} (\bbeta, \bgamma) &= \left\{\sum_{k=1}^{K} \sum_{i=1}^{n_k} \log P(Y_{ki}|\bX_{ki})+ D_{ki} (\bgamma_1^T \bX_{ki}+\gamma_2 Y_{ki}) \right\} + \sum_{i=1}^{K} n_{1k} \log{\xi_k} \\
&~~~ - \sum_{k=1}^{K} \sum_{i=1}^{n_k} \log \left\{ 1- \frac{n_{1k}}{n_{k}} + \frac{n_{1k}}{n_{k}} \xi_k \int_{\mathcal{Y}} P(y|\bx_{ik}) e^{\bgamma_1^T \bX_{ki}+\gamma_2 y} \, dy \right\}.
\end{split}
\]

We next consider the scenario of known disease rate. Suppose that the disease rate in the $k$th stratum $\xi_{0k} = P(D=1|Z=k)$ is known for $k=1,...,K$. The likelihood function reduces to 
\[
L(\btheta) = \prod_{k=1}^{K} \prod_{i=1}^{n_k}  \frac{p_{ki}P(Y_{ki}|\bX_{ki})  e^{D_{ki}(\gamma_{0k} + \bgamma_1^T \bX_{ki}+\gamma_2 Y_{ki})}}{1+ e^{\gamma_{0k} + \bgamma_1^T \bX_{ki}+\gamma_2 Y_{ki}} }.
\]
Besides the original constraints on $p_{ki}$s, however, there are additional constraints on $\btheta$ such that 
  \[
  \sum_{i=1}^{n_k} p_{ki} \int_{\mathcal{Y}} P(y|\bX_{ki}) \frac{ e^{\gamma_{0k} +\bgamma_1^T \bX_{ki}+\gamma_2 y}}{1+e^{\gamma_{0k} +\bgamma_1^T \bX_{ki}+\gamma_2 y}}dy =\xi_{0k}, ~~k=1,...,K.
  \]

Using the Lagrange multiplier, we maximize
\[
\begin{split}
g(\btheta, \lambda_1,...,\lambda_K, \tilde{\lambda}_1,...,\tilde{\lambda}_K) = & l(\btheta) - \sum_{k=1}^K \tilde{\lambda}_k \left(\sum_{i=1}^{n_k} p_{ki}-1 \right)\\
& - \sum_{k=1}^K \lambda_k \left\{\sum_{i=1}^{n_k} p_{ki} \int_{\mathcal{Y}} P(y|\bX_{ki}) \frac{ e^{\gamma_{0k} +\gamma_1 \bX_{ki}+\gamma_2 y}}{1+e^{\gamma_{0k} +\gamma_1 \bX_{ki}+\gamma_2 y}}dy -\xi_{0k}\right\}.\\
\end{split}
\]
We calculate the first derivative of $g(\btheta, \lambda_1,...,\lambda_K, \tilde{\lambda}_1,...,\tilde{\lambda}_K)$ with respect to $p_{ki}$ and set it to 0,
\[
\frac{\partial g(\btheta, \lambda_1,...,\lambda_K, \tilde{\lambda}_1,...,\tilde{\lambda}_K) }{\partial p_{ki}} = \frac{1}{p_{ki}} - \lambda_k \int_{\mathcal{Y}} P(y|\bX_{ki}) \frac{e^{\gamma_{0k}+\bgamma_1^T \bX_{ki}+\gamma_2 y}}{1+e^{\gamma_{0k}+\bgamma_1^T \bX_{ki}+\gamma_2 y} }dy  - \tilde{\lambda}_k=0.
\]
Multiplying the above equation by \(p_{ki}\) and summing it over $i=1,...,n_k$, we obtain 
\(\lambda_k \xi_{0k} + \tilde{\lambda}_k = n_k\) for \(k=1, ..., K\). 

Plugging \(\tilde{\lambda}_k = n_k - \xi_{0k} \lambda_k\) into the above equation, we can show that
\[
p_{ki} = \left\{ \lambda_k \int_{\mathcal{Y}} P(y|\bX_{ki}) \frac{e^{\gamma_{0k}+\bgamma_1^T \bX_{ki}+\gamma_2 y}}{1+e^{\gamma_{0k}+\bgamma_1^T \bX_{ki}+\gamma_2 y} }dy + n_k - \xi_{0k} \lambda_k \right\}^{-1}.
\]
Given $\bbeta$ and $\bgamma$, 
 \(\lambda_k\) can be determined by solving the following equation
 \[
\sum_{i=1}^{n_k} \left\{ \lambda_k \int_{\mathcal{Y}} P(y|\bX_{ki}) \frac{e^{\gamma_{0k}+\bgamma_1^T \bX_{ki}+\gamma_2 y}}{1+e^{\gamma_{0k}+\bgamma_1^T \bX_{ki}+\gamma_2 y} }dy + n_k - \xi_{0k} \lambda_k \right\}^{-1}=1.
\]
We denote the solution for $\lambda_k$ as $\lambda_k(\bbeta,\bgamma)$. The profile log-likelihood for $(\bbeta,\bgamma)$ is then given by 
\[
\begin{split}
l_{p} (\bbeta, \bgamma) &= \sum_{k=1}^{K} \sum_{i=1}^{n_k} \{\log P(Y_{ki}|\bX_{ki}) + D_{ki} (\gamma_{0k}+\bgamma_1^T \bX_{ki}+\gamma_2 Y_{ki})-\log(1+e^{\gamma_{0k}+\bgamma_1^T \bX_{ki}+\gamma_2 Y_{ki}}) \} \\
&~~~ - \sum_{k=1}^{K} \sum_{i=1}^{n_k} \log \left\{ \lambda_k(\bbeta,\bgamma) \int_{\mathcal{Y}} P(y|\bX_{ki}) \frac{e^{\gamma_{0k}+\bgamma_1^T \bX_{ki}+\gamma_2 y}}{1+e^{\gamma_{0k}+\bgamma_1^T \bX_{ki}+\gamma_2 y}} \, dy +n_k- \xi_{0k} \lambda_k(\bbeta,\bgamma)\right\}.\\
\end{split}
\]

We finally consider the scenario where the disease rate is unknown and not rare. 
The likelihood function is
\[
\begin{split}
L(\btheta) &= \prod_{k=1}^{K} \prod_{i=1}^{n_k} \left\{ \frac{p_{ki} P(Y_{ki}|\bX_{ki}) e^{\gamma_{0k} + \bgamma_1^T \bX_{ki} + \gamma_2 Y_{ki}}/(1+ e^{\gamma_{0k} + \bgamma_1^T \bX_{ki} + \gamma_2 Y_{ki}})}{P(D_{ki}=1)} \right\}^{D_{ki}} \\
&~~~ \times \prod_{k=1}^{K} \prod_{i=1}^{n_k} \left\{ \frac{p_{ki} P(Y_{ki}|\bX_{ki}) /(1+ e^{\gamma_{0k} + \bgamma_1^T \bX_{ki} + \gamma_2 Y_{ki}})}{P(D_{ki}=0)} \right\}^{1-D_{ki}}.
\end{split}
\]
Again, $p_{ki}$s satisfy the constraints 
\(\sum_{i=1}^{n_k} p_{ki} =1\) for \(k=1, ..., K.\) 
Using the Lagrange multiplier, we maximize
\[
g(\btheta, \lambda_1,...,\lambda_K) = l(\btheta) - \sum_{k=1}^K \lambda_k \left(\sum_{i=1}^{n_k} p_{ki}-1 \right).
\]
We calculate the first derivative of $g(\btheta, \lambda_1,...,\lambda_K)$ with respect to $p_{ki}$ and set it to 0,
\[
\begin{split}
\frac{\partial g(\btheta, \lambda_1,...,\lambda_K) }{\partial p_{ki}} &= \frac{1}{p_{ki}} - \frac{n_{1k}}{P(D_{ki}=1)} \int_{\mathcal{Y}} P(y|\bX_{ki}) \frac{e^{\gamma_{0k} + \bgamma_1^T \bX_{ki} + \gamma_2 y}}{1+e^{\gamma_{0k} + \bgamma_1^T \bX_{ki} + \gamma_2 y}} \,dy \\
&~~~ - \frac{n_{0k}}{P(D_{ki}=0)} \int_{\mathcal{Y}} P(y|\bX_{ki}) \frac{1}{1+e^{\gamma_{0k} + \bgamma_1^T \bX_{ki} + \gamma_2 y}} \, dy- \lambda_k=0.
\end{split}
\]
Multiplying the above equation by \(p_{ki}\) and summing it over $i=1,...,n_k$, we obtain 
\(\lambda_k = 0\) for \(k=1, ..., K\).\\
Plugging \(\lambda_k = 0\) into the above equation, we can show that
\[
p_{ki} = \left\{ \frac{n_{1k}}{\xi_k} \int_{\mathcal{Y}} P(y|\bX_{ki}) \frac{e^{\gamma_{0k}+\bgamma_1^T \bX_{ki} + \gamma_2 y}}{1+e^{\gamma_{0k}+\bgamma_1^T \bX_{ki} + \gamma_2 y}} \, dy + \frac{n_{0k}}{1-\xi_k} \int_{\mathcal{Y}} P(y|\bX_{ki}) \frac{1}{1+e^{\gamma_{0k}+\bgamma_1^T \bX_{ki} + \gamma_2 y}} \, dy\right\}^{-1},
\]
where the disease rate in the $k$th stratum \(\xi_k = P(D_{ik}=1|Z=k)\) is unknown for $k=1,...,K$. Write $\bxi=(\xi_1,...,\xi_K)$. 
The profile log-likelihood for $(\bbeta,\bgamma,\bxi)$ is then given by 
\[
\begin{split}
l_{p} (\bbeta,\bgamma,\bxi) &= \sum_{k=1}^{K} \sum_{i=1}^{n_k} \{ D_{ki} (\gamma_{0k}+\bgamma_1^T \bX_{ki}+\gamma_2 Y_{ki})-\log(1+e^{\gamma_{0k}+\bgamma_1^T \bX_{ki} + \gamma_2 Y_{ki}}) + P(Y_{ki}|\bX_{ki}) \} \\
&~~~ -\sum_{i=1}^{K} n_{1k} \xi_k -\sum_{i=1}^{K} n_{0k} (1-\xi_{k}) - \sum_{k=1}^{K} \sum_{i=1}^{n_k} \log \bigg\{ \frac{n_{1k}}{\xi_k} \int_{\mathcal{Y}} P(y|\bX_{ki}) \frac{e^{\gamma_{0k}+\bgamma_1^T \bX_{ki}+\gamma_2 y}}{1+e^{\gamma_{0k}+\bgamma_1^T \bX_{ki} + \gamma_2 y}} \, dy \\
&~~~ + \frac{n_{0k}}{1-\xi_k} \int_{\mathcal{Y}} P(y|\bX_{ki}) \frac{1}{1+e^{\gamma_{0k}+\bgamma_1^T \bX_{ki} + \gamma_2 y}} \, dy\bigg\}.
\end{split}
\]


Once we obtain the profile log-likelihood function of $(\bbeta, \bgamma)$ or $(\bbeta,\bgamma,\bxi)$, we use the BFGS algorithm implemented in the {\it optim} R package to get the maximum likelihood estimator of $(\bbeta,\bgamma)$ or $(\bbeta,\bgamma,\bxi)$, denoted by $(\widehat{\bbeta}_n, \widehat{\bgamma}_n)$ or $(\widehat{\bbeta}_n, \widehat{\bgamma}_n,\widehat{\bxi}_n)$. We also want to note that the identifiability of $(\bbeta,\bgamma,\bp)$ was shown by \cite{lin_proper_2009}, but it is not clear that the unknown disease rates are identifiable in the last scenario. When disease rates are not identifiable or only weakly identifiable due to a flat profile likelihood function of $\xi_k$, the algorithms may fail to converge and estimates may be unstable. Even though PM3 achieved satisfactory performance in our simulations, we recommend obtaining information on disease rates and using PM2, rather than estimating disease rates using PM3.

\subsection*{A.2.\enspace Connection with Empirical Likelihood Approach}\label{empl}
\vspace*{12pt}

Our problem of maximizing a likelihood function under some constraints (e.g., known disease rates) is very similar to the problems solved by the empirical likelihood approach \citep{owen2001empirical,qin1994empirical}. Typically, the profile empirical likelihood function is
\[
\mathcal{R}(\bmu)
= \max \left\{
    \prod_{i=1}^n w_i
    \,\Big|\,
    \sum_{i=1}^n w_i \bX_i = \bmu,\;
    w_i \ge 0,\;
    \sum_{i=1}^n w_i = 1
\right\},
\]
where $\bX_i ~(i=1,...,n)$ and $\bmu$ are $d\times 1$ vectors. 
This is an example of convex duality. As described by \cite{owen2001empirical}, we can convert this maximization over $n$ variables $w_i$ subject to $d+1$ equality constraints into a minimization over $d$ variables $\blambda$,
\[
\log R(F)
= - \sum_{i=1}^n \log\big\{ 1 + \blambda^T(\bX_i - \bmu) \big\},
\]
where $F$ is the distribution function of $\bX$. In our implementation, we create a routine to solve for $\blambda$ based on this duality.
In addition, the likelihood functions we derived above are subject to the constraints that the arguments of the log functions are positive. Instead of explicitly imposing these constraints, we replace the log function with a pseudo-logarithm function proposed by \cite{owen2001empirical},
\[
\log_\star(z) =
\begin{cases}
\log(z), & \text{if } z \ge 1/n,\\[6pt]
\log(1/n) - 1.5 + 2nz - (nz)^2/2, & \text{if } z < 1/n.
\end{cases}
\]
Now we can maximize the likelihood function without the inequality constraints, which significantly simplifies the optimization problem. 

\subsection*{A.3.\enspace Asymptotic Properties}\label{asymptotic-properties}
\vspace*{12pt}

The asymptotic properties of $(\widehat{\bbeta}_n,\widehat{\bgamma}_n)$ under case-control sampling
have been established \citep{lin_likelihood-based_2006}. Using arguments similar to those developed in earlier work \citep{lin_likelihood-based_2006,lin_proper_2009}, we can show that the
maximum likelihood estimators are consistent and asymptotically normal.
In addition, the limiting covariance matrix attains the efficiency bound
and can be consistently estimated by the inverse of the negative Hessian
matrix of the profile log-likelihood function \citep{murphy2000profile}.

Below, we sketch the proof of the asymptotic properties of the proposed estimator. Now consider a matched case-control design with \(K\) strata, indexed by $k=1,\dots,K$. In stratum
\(k\):
\begin{enumerate}
\item 
  The number of strata K is fixed.
\item
  Sampling is independent across strata.
\item
  A case-control sample of size \(n_k\) is drawn independently.
\item
  The parameters $(\bbeta_1,\bgamma_1,\gamma_2)$ are the same across strata. The other nuisance parameters can vary among strata.
\end{enumerate}

Note that $\bbeta$ and $\bgamma$ were defined in Section 2 and  include stratum-specific parameters $\beta_{0k}$ and $\gamma_{0k}$. With a slight abuse of notation, we let $\btheta=(\bbeta,\bgamma)$ denote all the unknown parameters and $\btheta_0$ the true parameter values. The stratum-specific parameters and their true values are denoted by $\btheta_k$ and $\btheta_{0k}$ for stratum $k$. Define the stratified estimator \(\widehat\btheta_{nk}\) as the
maximizer of the profile log-likelihood function
\[
l_{p,n} (\theta) = \sum_{k=1}^{K} l_{p,n_k,k}(\btheta_k),~~\text{where}~~ n=\sum_{k=1}^{K} n_k, ~~\text{and}~~ n_k/n \rightarrow \pi_k \in (0,1).
\]

Our setup can be viewed as an application of empirical likelihood for regression models. Asymptotic properties of EL-based estimators have been established in the literature\citep{owen2001empirical,chen2009review,wu2011empirical}. Thus, our proposed estimators inherit the same efficiency and asymptotic normality as standard empirical likelihood estimators developed for regression models.

\subsubsection*{A.3.1.\enspace Consistency}\label{consistency}

The consistency of the unstratified estimator (i.e., pooled, single-stratum) is established in Theorem 2 of \cite{lin_likelihood-based_2006} and applies to $\hat{\btheta}_{nk}$. Since $\btheta_{0k} = (\bgamma_{0k},\bbeta_0)$ uniquely maximizes the stratum log likelihood, $\btheta_{0}=(\bgamma_{01},\dots,\bgamma_{0K},\bbeta_0)$ uniquely maximizes the overall log likelihood. So $\hat{\btheta}_n$ is consistent for $\btheta_0$.

\subsection*{A.3.2.\enspace Asymptotic Normality}\label{asymptotic-normality}

Asymptotic normality for the unstratified estimators is also proved. The derivation is similar to that of theorem 1.2 of \cite{murphy_semiparametric_2001}. Linearization of the score function
around the true parameter value shows that it converges weakly to a
Gaussian process, and that the operator is invertible. Then by
theorem 3.3.1 of \cite{vaart_weak_1996}, asymptotic normality follows.

\subsection*{A.3.3.\enspace Asymptotic Efficiency}\label{asymptotic-efficiency}

The asymptotic efficiency results for the unstratified estimators follow from proposition 3.3.1 of \cite{bickel_efficient_1993} that the limiting covariance matrix attains the semiparametric efficiency bound.

\bibliographystyle{agsm}
\bibliography{MCC}

@book{bickel_efficient_1993,
  title={Efficient and adaptive estimation for semiparametric models},
  author={Bickel, Peter J and Klaassen, Chris AJ and Bickel, Peter J and Ritov, Ya’acov and Klaassen, J and Wellner, Jon A and Ritov, YA'Acov},
  volume={4},
  year={1993},
  publisher={Springer}
}

@article{murphy2000profile,
  title={On profile likelihood},
  author={Murphy, Susan A and Van der Vaart, Aad W},
  journal={Journal of the American Statistical Association},
  volume={95},
  number={450},
  pages={449--465},
  year={2000},
  publisher={Taylor \& Francis}
}

@article{lin_likelihood-based_2006,
  title={Likelihood-based inference on haplotype effects in genetic association studies},
author={Lin, DY and Zeng, D}, 
  journal={Journal of the American Statistical Association},
  volume={101},
  number={473},
  pages={89--104},
  year={2006},
  publisher={Taylor \& Francis}
}

@article{murphy_semiparametric_2001,
  title={Semiparametric mixtures in case-control studies},
  author={Murphy, S A and Van der Vaart, A W},
  journal={Journal of Multivariate Analysis},
  volume={79},
  number={1},
  pages={1--32},
  year={2001},
  publisher={Elsevier}
}

@book{vaart_weak_1996,
	location = {New York},
	title = {Weak convergence and empirical processes},
	isbn = {0-387-94640-3},
	series = {Springer series in statistics},
	publisher = {Springer},
	author = {Vaart, A. W. van der. and Wellner, Jon A.},
	year = {1996},
}

@article{lin_proper_2009,
  title={Proper analysis of secondary phenotype data in case-control association studies},
  author={Lin, DY and Zeng, D},
  journal={Genetic Epidemiology},
  volume={33},
  number={3},
  pages={256--265},
  year={2009},
  publisher={Wiley Online Library}
}

@book{lash_modern_2021,
  title     = {Modern Epidemiology},
  author    = {Lash, Timothy L. and VanderWeele, Tyler J. and Haneuse, Sebastien and Rothman, Kenneth J.},
  edition   = {4},
  year      = {2021},
  publisher = {Wolters Kluwer},
  address   = {Philadelphia, PA},
  isbn      = {9781451193282}
}

@article{tchetgen_tchetgen_general_2014,
  title={A general regression framework for a secondary outcome in case--control studies},
  author={Tchetgen Tchetgen, Eric J},
  journal={Biostatistics},
  volume={15},
  number={1},
  pages={117--128},
  year={2014},
  publisher={Oxford University Press}
}

@article{gail_likelihood_1981,
  title={Likelihood calculations for matched case-control studies and survival studies with tied death times},
  author={Gail, Mitchell H and Lubin, Jay H and Rubinstein, Lawrence V},
  journal={Biometrika},
  pages={703--707},
    volume={68},
  year={1981},
  publisher={JSTOR}
}

@book{therneau_package_2023,
	title = {A Package for Survival Analysis in R},
	url = {https://CRAN.R-project.org/package=survival},
	author = {Therneau, Terry M.},
	year = {2023},
}

@incollection{diao_analysis_2018,
  title={Analysis of Secondary Phenotype Data under Case-Control Designs},
  author={Diao, Guoqing and Zeng, Donglin and Lin, Dan-Yu},
  booktitle={Handbook of Statistical Methods for Case-Control Studies},
  pages={515--528},
  year={2018},
  publisher={Chapman and Hall/CRC}
}

@article{he2012gaussian,
  title={A Gaussian copula approach for the analysis of secondary phenotypes in case--control genetic association studies},
  author={He, Jing and Li, Hongzhe and Edmondson, Andrew C and Rader, Daniel J and Li, Mingyao},
  journal={Biostatistics},
  volume={13},
  number={3},
  pages={497--508},
  year={2012},
  publisher={Oxford University Press}
}

@article{ma2016semiparametric,
  title={Semiparametric estimation in the secondary analysis of case--control studies},
  author={Ma, Yanyuan and Carroll, Raymond J},
  journal={Journal of the Royal Statistical Society Series B: Statistical Methodology},
  volume={78},
  number={1},
  pages={127--151},
  year={2016},
  publisher={Oxford University Press}
}

@article{richardson2007analyses,
  title={Analyses of case--control data for additional outcomes},
  author={Richardson, David B and Rzehak, Peter and Klenk, Jochen and Weiland, Stephan K},
  journal={Epidemiology},
  volume={18},
  number={4},
  pages={441--445},
  year={2007},
  publisher={LWW}
}

@article{monsees2009genome,
  title={Genome-wide association scans for secondary traits using case-control samples},
  author={Monsees, Genevieve M and Tamimi, Rulla M and Kraft, Peter},
  journal={Genetic Epidemiology},
  volume={33},
  number={8},
  pages={717--728},
  year={2009},
  publisher={Wiley Online Library}
}

@article{ghosh2013unified,
  title={Unified analysis of secondary traits in case--control association studies},
  author={Ghosh, Arpita and Wright, Fred A and Zou, Fei},
  journal={Journal of the American Statistical Association},
  volume={108},
  number={502},
  pages={566--576},
  year={2013},
  publisher={Taylor \& Francis}
}

@article{wei2013robust,
  title={Robust estimation for homoscedastic regression in the secondary analysis of case--control data},
  author={Wei, Jiawei and Carroll, Raymond J and M{\"u}ller, Ursula U and Keilegom, Ingrid Van and Chatterjee, Nilanjan},
  journal={Journal of the Royal Statistical Society Series B: Statistical Methodology},
  volume={75},
  number={1},
  pages={185--206},
  year={2013},
  publisher={Oxford University Press}
}

@article{song2016general,
  title={A general and robust framework for secondary traits analysis},
  author={Song, Xiaoyu and Ionita-Laza, Iuliana and Liu, Mengling and Reibman, Joan and Wei, Ying},
  journal={Genetics},
  volume={202},
  number={4},
  pages={1329--1343},
  year={2016},
  publisher={Oxford University Press}
}

@misc{diabetes_130,
  author       = {Clore, John and Cios, Krzysztof and DeShazo, Jon and Strack, Beata},
  title        = {{Diabetes 130-US Hospitals for Years 1999-2008}},
  year         = {2014},
  howpublished = {UCI Machine Learning Repository},
  note         = {{DOI}: https://doi.org/10.24432/C5230J}
}

@book{owen2001empirical,
  title={Empirical likelihood},
  author={Owen, Art B},
  year={2001},
  publisher={Chapman and Hall/CRC}
}

@article{qin1994empirical,
  title={Empirical likelihood and general estimating equations},
  author={Qin, Jin and Lawless, Jerry},
  journal={The Annals of Statistics},
  volume={22},
  number={1},
  pages={300--325},
  year={1994},
  publisher={Institute of Mathematical Statistics}
}

@article{zhou2020analysis,
  title={Analysis of secondary phenotypes in multigroup association studies},
  author={Zhou, Fan and Zhou, Haibo and Li, Tengfei and Zhu, Hongtu},
  journal={Biometrics},
  volume={76},
  number={2},
  pages={606--618},
  year={2020},
  publisher={Oxford University Press}
}

@article{pan2018secondary,
  title={Secondary outcome analysis for data from an outcome-dependent sampling design},
  author={Pan, Yinghao and Cai, Jianwen and Longnecker, Matthew P and Zhou, Haibo},
  journal={Statistics in Medicine},
  volume={37},
  number={15},
  pages={2321--2337},
  year={2018},
  publisher={Wiley Online Library}
}

@article{jordan2019communication,
  title={Communication-efficient distributed statistical inference},
  author={Jordan, Michael I and Lee, Jason D and Yang, Yun},
  journal={Journal of the American Statistical Association},
  year={2019},
    volume={114},
    pages={668--681},
  publisher={Taylor \& Francis}
}

@article{chen2014split,
  title={A split-and-conquer approach for analysis of extraordinarily large data},
  author={Chen, Xueying and Xie, Min-ge},
  journal={Statistica Sinica},
  pages={1655--1684},
  year={2014},
  publisher={JSTOR}
}

@article{zeng2015random,
  title={On random-effects meta-analysis},
  author={Zeng, D and Lin, DY},
  journal={Biometrika},
  volume={102},
  number={2},
  pages={281--294},
  year={2015},
  publisher={Oxford University Press}
}

@article{lin2010relative,
  title={On the relative efficiency of using summary statistics versus individual-level data in meta-analysis},
  author={Lin, DY and Zeng, D}, 
  journal={Biometrika},
  volume={97},
  number={2},
  pages={321--332},
  year={2010},
  publisher={Oxford University Press}
}

@article{battey2018distributed,
  author = {Heather Battey and Jianqing Fan and Han Liu and Junwei Lu and Ziwei Zhu},
  title = {{Distributed testing and estimation under sparse high dimensional models}},
  volume = {46},
  journal = {The Annals of Statistics},
  number = {3},
  publisher = {Institute of Mathematical Statistics},
  pages = {1352 -- 1382},
  year = {2018},
  doi = {10.1214/17-AOS1587},
  URL = {https://doi.org/10.1214/17-AOS1587}
}

@article{chen2009review,
  title={A review on empirical likelihood methods for regression},
  author={Chen, Song Xi and Van Keilegom, Ingrid},
  journal={Test},
  volume={18},
  number={3},
  pages={415--447},
  year={2009},
  publisher={Springer}
}

@article{wu2011empirical,
  title={An empirical likelihood approach to nonparametric covariate adjustment in randomized clinical trials},
  author={Wu, Xiaoru and Ying, Zhiliang},
  journal={arXiv preprint arXiv:1108.0484},
  year={2011}
}

@article{slud2018combining,
  title={Combining estimators of a common parameter across samples},
  author={Slud, Eric and Vonta, Ilia and Kagan, Abram},
  journal={Statistical Theory and Related Fields},
  volume={2},
  number={2},
  pages={158--171},
  year={2018},
  publisher={Taylor \& Francis}
}

@article{chen2020relative,
  title={Relative efficiency of using summary versus individual data in random-effects meta-analysis},
  author={Chen, Ding-Geng and Liu, Dungang and Min, Xiaoyi and Zhang, Heping},
  journal={Biometrics},
  volume={76},
  number={4},
  pages={1319--1329},
  year={2020},
  publisher={Oxford University Press}
}

@inproceedings{finn2017model,
  title={Model-agnostic meta-learning for fast adaptation of deep networks},
  author={Finn, Chelsea and Abbeel, Pieter and Levine, Sergey},
  booktitle={International conference on machine learning},
  pages={1126--1135},
  year={2017},
  organization={PMLR}
}

@article{zhuang2020comprehensive,
  title={A comprehensive survey on transfer learning},
  author={Zhuang, Fuzhen and Qi, Zhiyuan and Duan, Keyu and Xi, Dongbo and Zhu, Yongchun and Zhu, Hengshu and Xiong, Hui and He, Qing},
  journal={Proceedings of the IEEE},
  volume={109},
  number={1},
  pages={43--76},
  year={2020},
  publisher={Ieee}
}

@book{schlesselman1982case,
  title={Case-control studies: design, conduct, analysis},
  author={Schlesselman, James J},
  volume={2},
  year={1982},
  publisher={Oxford university press}
}

@book{breslow1980statistical,
  title={Statistical methods in cancer research. Vol. 1. The analysis of case-control studies.},
  author={Breslow, Norman E and Day, NE},
  number={32},
  publisher={IARC Scientific Publications},
  year={1980}
}

@article{prentice1979logistic,
  title={Logistic disease incidence models and case-control studies},
  author={Prentice, Ross L and Pyke, Ronald},
  journal={Biometrika},
  volume={66},
  number={3},
  pages={403--411},
  year={1979},
  publisher={Oxford University Press}
}

@article{thomas2010methods,
  title={Methods for investigating gene-environment interactions in candidate pathway and genome-wide association studies},
  author={Thomas, Duncan},
  journal={Annual review of public health},
  volume={31},
  number={1},
  pages={21--36},
  year={2010},
  publisher={Annual Reviews}
}

@article{balding2006tutorial,
  title={A tutorial on statistical methods for population association studies},
  author={Balding, David J},
  journal={Nature reviews genetics},
  volume={7},
  number={10},
  pages={781--791},
  year={2006},
  publisher={Nature Publishing Group UK London}
}

@article{pearce2016analysis,
  title={Analysis of matched case-control studies},
  author={Pearce, Neil},
  journal={bmj},
  volume={352},
  year={2016},
  publisher={British Medical Journal Publishing Group}
}

@book{rothman2008modern,
  title={Modern epidemiology},
  author={Rothman, Kenneth J and Greenland, Sander and Lash, Timothy L and others},
  volume={3},
  year={2008},
  publisher={Wolters Kluwer Health/Lippincott Williams \& Wilkins Philadelphia}
}

@article{mantel1959statistical,
  title={Statistical aspects of the analysis of data from retrospective studies of disease},
  author={Mantel, Nathan and Haenszel, William},
  journal={Journal of the national cancer institute},
  volume={22},
  number={4},
  pages={719--748},
  year={1959},
  publisher={Oxford University Press}
}

@article{cox1972regression,
  title={Regression models and life-tables},
  author={Cox, David R},
  journal={Journal of the Royal Statistical Society: Series B (Methodological)},
  volume={34},
  number={2},
  pages={187--202},
  year={1972},
  publisher={Wiley Online Library}
}
\end{document}